\documentclass[natbib]{svjour3}                     
\smartqed           
\usepackage{graphicx}
\usepackage{lscape}

\begin{document}
\title{Large Amplitude Oscillations in Prominences}
\titlerunning{Large Amplitude Oscillations in Prominences}
\author{D. Tripathi \and H. Isobe \and R. Jain}
\authorrunning{Tripathi, Isobe and Jain}
\institute{D. Tripathi \at Department of Applied Mathematics and Theoretical Physics, University of
Cambridge, Wilberforce Road, Cambridge CB3 0WA, UK \\
                  Tel.: +44-1323-337916\\
                  \email{d.tripathi@damtp.cam.ac.uk}
               \and
               H. Isobe \at Unit of Synergetic Studies for Space, Kyoto
                University, Yamashina, Kyoto 607-8471, Japan \\
                \and
                R. Jain \at School of Mathematics and Statistics, University of
                Sheffield, Western Bank, Sheffield S3 7RH, UK}
    \date{Received: date / Accepted: date}
    \maketitle

\begin{abstract}
Since the first reports of oscillations in prominences in 1930s
there have been major theoretical and observational advances to
understand the nature of these oscillatory phenomena leading to a
whole new field of so called "prominence seismology". There are
two types of oscillatory phenomena observed in prominences; "small
amplitude oscillations" (~2-3 km~s$^{-1}$) which are quite common
and "large amplitude oscillations" ($>$20 km~s$^{-1}$) for which
observations are scarce.  Large amplitude oscillations have been
found as "winking filament" in H$\alpha$ as well as motion in the
sky plane in H$\alpha$, EUV, micro-wave and He 10830 observations.
Historically, it was suggested that the large amplitude
oscillations in prominences were triggered by disturbances such as
fast-mode MHD waves (Moreton wave) produced by remote flares.
Recent observations show, in addition, that near-by flares or jets
can also create such large amplitude oscillations in prominences.
Large amplitude oscillations, which are observed both in
transverse as well as longitudinal direction, have a range of
periods varying from tens of minutes to a couple of hours. Using
the observed period of oscillation and simple theoretical models,
the obtained magnetic field in prominences has shown quite a good
agreement with directly measured one and therefore, justifies
prominences seismology as a powerful diagnostic tool. On rare
occasions, when the large amplitude oscillations have been
observed before or during the eruption, the oscillations may be
applied to diagnose the stability and the eruption mechanism. Here
we review the recent developments and understanding in the
observational properties of large amplitude oscillations and their
trigger mechanisms and stability in the context of prominence
seismology.

\end{abstract}

\section{Introduction}\label{intro}

The existence of oscillatory motion in prominences has been known since 1930s \citep[see e.g., ][for a historical review]{hyder_1966}.  However, The first systematic investigation of prominence oscillations were performed by \cite{ramsey_smith_1966}. The filament oscillations reported by \cite{ramsey_smith_1966} were believed to be excited due to a distant flare. Filaments and prominences are the same entities with filaments seen in absorption against the solar disk and the prominences over the solar limb in emission. We will use the terms filament and prominence interchangeably throughout the paper.

The prominence oscillations have been broadly classified into two groups based on their observed velocity amplitudes, namely, the 'large amplitude oscillations' and the 'small amplitude oscillations' \citep[see e.g.,][for excellent reviews]{oliver_review, oliver_2009}. The large amplitude oscillations occur when the prominence exhibits large displacements, from a few thousands of kilometers up to 4 $\times$10$^4$~km \citep[][]{jing_2006} from its equilibrium position and the prominence as a whole oscillates with velocity amplitude of the order of 20~km~s$^{-1}$ or more. On the other hand the small amplitude oscillations are with velocity amplitudes of the order of 2$-$3~km~s$^{-1}$ or sometimes even less. In the last few decades, there have been numerous observations reporting small amplitude oscillations and their theoretical interpretation \citep[see e.g.,][]{ stephane, oliver_review, oliver_2009}. However, observations of large amplitude oscillations are scarce. The purpose of this article is to provide an up to date overview of the observations of large-amplitude oscillations reported in literature and discuss their importance in the context of prominence seismology.

The rest of the paper is organized as follows. In the next section (section~\ref{obs}) we summarize the observations of large amplitude oscillations to date. In section ~\ref{params} we provide an up-to-date summary of the physical parameters of large-amplitude oscillation events followed by a discussion on their trigger mechanisms in section~\ref{trigger}. In section~\ref{seismology} the application of prominence seismology technique to large amplitude oscillations with a view to determine the internal magnetic field is discussed; its application to the eruption of oscillating prominences is discussed in section~\ref{eruption}. Finally, a brief summary and a general discussion is presented in section~\ref{summary}.

\section{Observations of large amplitude oscillations in prominences \label{obs}}

\begin{figure}
\centering
\includegraphics[width=0.9\textwidth]{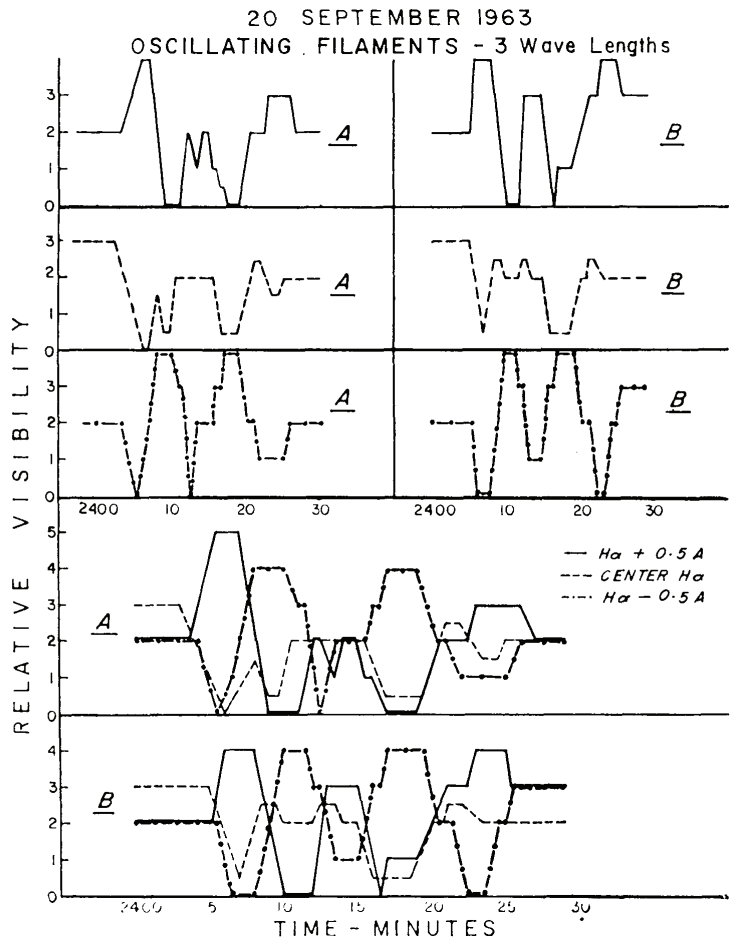}
\caption{Relative visibility of two segments namely A and B of a filament seen in H$\alpha$ line center and H$\alpha$$\pm$0.5~{\AA}. In the plot, solid lines correspond to relative visibility in H$\alpha$+0.5~{\AA}, dashed line for H$\alpha$ line center and dashed-dotted line correspond to H$\alpha$-0.5~{\AA}. Figure is adopted from \cite{ramsey_smith_1966}.\label{rs}}
\end{figure}

\begin{figure}
\centering
\includegraphics[width=1.0\textwidth]{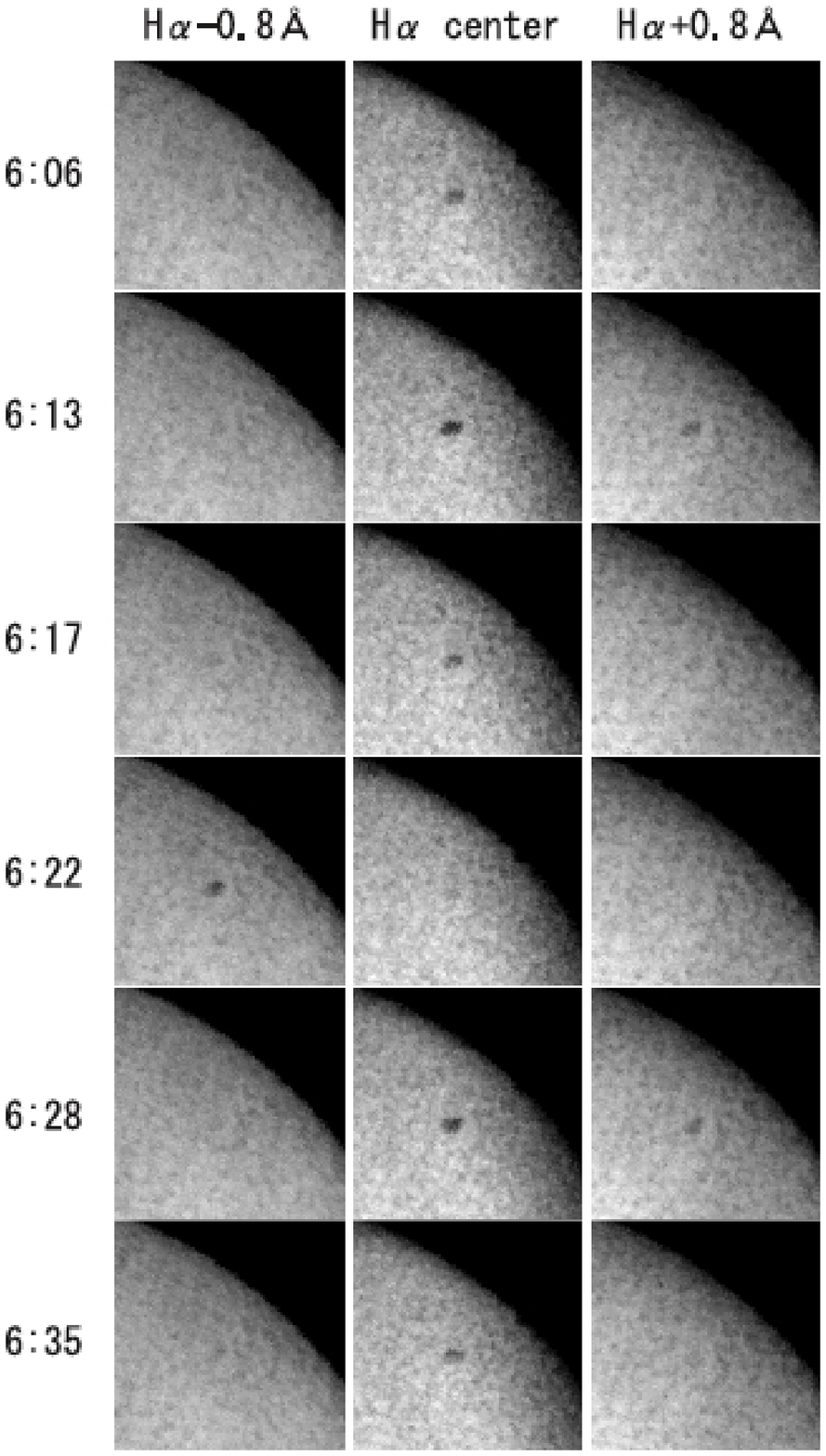}
\caption{A winking filament taken by the Flare Monitoring Telescope \citep{kurokawa_1995}. H$\alpha$ center
and $\pm$ 0.8~{\AA}. Figure is adopted from \cite{eto_2002}. \label{eto}}
\end{figure}

There have been a handful of observations reporting on the large amplitude oscillations in prominences. The first systematic study of large amplitude oscillations was performed by \cite{ramsey_smith_1966}, which were observed using H$\alpha$ spectrograms. The filament was seen in H$\alpha$ blue wing, followed by H$\alpha$ line center and then in H$\alpha$ red wing (see Fig.~\ref{rs}). Due to the appearance and disappearance of the filament in the H$\alpha$ line center images, this phenomenon is also knows as 'winking filament'. Fig.~\ref{eto} shows an example of a winking filament studied by \cite{eto_2002}.  In the figure, the filament first becomes visible in the red wing and then in the blue wing, suggesting that the filament was first pushed downward. This is consistent with the interpretation that the oscillation was triggered by a fast-mode MHD shock wave originated from a distant flare (Moreton wave; \cite{moreton}). Since the speed of a fast-mode wave is relatively faster in the corona than in the chromosphere, the normal vector of the wave front points downward as the wave travels from the flare site \citep{uchida}, pushing the prominence downwards. See \cite{hyder_1966} for a historical overview of "winking filaments". Recently \cite{jing_2003, jing_2006} and \cite{vrsnak_2007} have also presented large amplitude oscillations in filaments as the plane-of-sky (POS) motion observed in H$\alpha$.

Large amplitude oscillations have also been observed in other wavelengths such as EUV and microwave as a motion in the POS. \cite{isobe_tripathi_2006} reported oscillations in an erupting prominence in the EUV images from SOHO/EIT. Later they discovered that the same oscillations were also observed in microwave (17~GHz) images \citep{isobeetal_2007} taken from Nobeyama Radio Heliograph. More recently, \cite{gilbert_2008} found a large-amplitude oscillation event in He~10830 images recorded at the Mauna Loa Solar Observatory (MLSO). We summarize all the large amplitude oscillations in prominences observed to date in Table~\ref{table}.


\section{Physical Parameters of large amplitude Oscillations \label{params}}

\subsection{Amplitude and Direction of Motion}  The amplitudes of oscillations, so called ``large amplitude oscillations'' are provided in Table~\ref{table}. As is clear from Table~\ref{table}, the amplitude of oscillations can be as large as $\approx$90~km~s$^{-1}$.

A number of the reported large amplitude oscillations exhibit transverse motion, i.e.,  perpendicular to the prominence/filament axis. In our picture, we have considered filaments as long cylindrical structures. Both nearly vertical and nearly horizontal motion have been observed. When the filament is seen on the disk, the line of sight (LOS) motion corresponds to nearly vertical motion of the filament, while on the limb, it corresponds to nearly horizontal motion.

The three-dimensional velocities can be obtained by combining the LOS velocities derived from the Doppler shift and the POS velocities derived by correlation tracking of specific features in a time series of images. However, it is often quite difficult to derive a reliable POS velocity from H$\alpha$ observation, since the visibility of filaments is significantly affected by the Doppler effect. \cite{isobe_tripathi_2006} used H$\alpha$ center and wing observations to derive LOS velocity using the method described by \cite{M_K_2003}, and the EUV images taken by the Extreme-ultraviolet Imaging Telescope \citep[EIT;][]{eit} to track the motion in the POS. They found that the POS speed was 5~km~s$^{-1}$ and the LOS velocity was $\sim$ 20~km~s$^{-1}$. Since the filament was near the southern-east limb (part of it was seen as a prominence in off-limb), the motion was nearly horizontal to the local solar photosphere.

\begin{figure}
\centering
\includegraphics[width=0.9\textwidth]{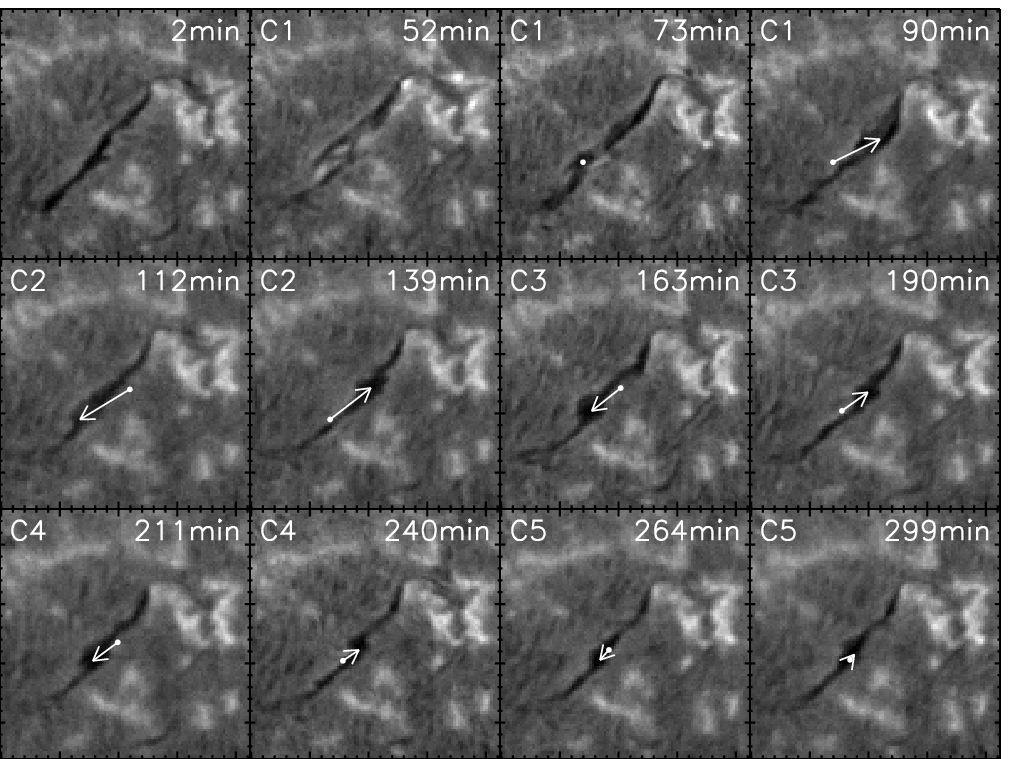}
\caption{Sequence of BBSO H$\alpha$ images for a subfield of solar disk. The observation time (in minutes after 17:00 UT) is given in each panel. The oscillation continued for five consecutive cycles (C1-C5). The arrows connect the determined displacement for consecutive images taken around the oscillation peaks for each of the five cycles. The decreasing lengths of the arrows directly reflect the instantaneous oscillation amplitudes and their damping. See \cite{vrsnak_2007} for an extended discussion. Figure is adopted from \cite{vrsnak_2007}.\label{vrsnak}}
\end{figure}

Recently, longitudinal oscillations, i.e., oscillations along the filament axis have been reported \citep{jing_2003, jing_2006, vrsnak_2007}. Fig.~\ref{vrsnak} shows such an example of longitudinal motion presented by \cite{vrsnak_2007}. In this observation, the plasma motion could be seen all along the filament, with the amplitude decreasing towards the legs of the filament, i.e., the amplitude of oscillations was largest in the middle part of the filament where the bulk feature is located (see top row, third panel in Fig.~\ref{vrsnak}). \cite{vrsnak_2007} interpreted this oscillation as a propagation of a longitudinal standing mode on a slinky spring which is fixed at both ends.

\subsection{Oscillation Period and Damping}

As summarized in Table 1, the observed periods of large amplitude oscillations reported in the literature ranges from 6 to 150 min. Out of 11 oscillation events analyzed by \cite{ramsey_smith_1966}, 4 events occurred in the same filament over a period of three days. Interestingly, all four oscillations in this filament had the same period of oscillation. \cite{ramsey_smith_1966} found that the filament had its own characteristic frequency of oscillation, which was independent of the size of the flare, the distance from the flare or the inferred wave velocity of the propagating disturbance. This is the basis of prominence seismology as discussed in section~\ref{seismology}.

In an erupting prominence associated with large amplitude oscillation in transverse horizontal direction, detected on 15-October-2002 in EUV observation, \cite{isobe_tripathi_2006} estimated a period of about 150 minutes which was later confirmed by \cite{pinter} with a more sophisticated data analysis technique namely wavelet-analysis. Wavelet analysis is a powerful tool to investigate localized variations of power within a time series \citep[see][]{torrence}. \cite{jing_2003, jing_2006, vrsnak_2007} reported longitudinal oscillations in different filaments recorded in H$\alpha$ observations with periods 150, 100, and 50 minutes respectively. Based on He~I~10830 observations, recently, \cite{gilbert_2008} reported oscillation in filament in transverse vertical direction with a period of 29 minutes. Based on the observations reported so far, it is difficult to see if the period of oscillation has any kind of dependence on the direction of the motion or the trigger of the oscillation (discussed in section~\ref{trigger}).


An extremely long period (8-27 hours) oscillation was reported by \cite{foullon} using wavelet analysis. However, since the oscillation was reported only in the intensity variation in SOHO/EIT 195~{\AA} images, it is not clear if the oscillation was accompanied by a large amplitude motion of the filament. So far, there have been no attempts to detect displacement and amplitude variation using H$\alpha$ filtergrams or by any other means for this particular event presented by \cite{foullon}. It would be interesting in the future to investigate if such an oscillation with ultra-long periods as presented by \cite{foullon, foullon_2009, pouget} has any clear association with large amplitude.

The prominence oscillation reported by \cite{isobe_tripathi_2006} and studied in detail by \cite{isobeetal_2007, pinter} erupted after a few cycles. Therefore, no damping in the oscillation was observed for this particular event. However, damping in large amplitude oscillation of prominences do exist and are reported by different authors \citep[see e.g.,][]{jing_2003, jing_2006, vrsnak_2007, gilbert_2008}. The observed damping time ranges from 2 to 6 times their periods. Although the damping mechanism for these oscillations has not been fully understood so far, it can be attributed to the energy losses by the emission of waves in to the ambient corona \citep[see e.g.,][]{k_k, miyagoshi} and/or to various dissipative processes \citep[see e.g., ][]{hyder_1966, nakariakov_1999, ofman_aschwanden_2002, ballai, verwichte_2004}. In addition, the mechanism at work in damping of large-amplitude oscillations could be similar to those for small-amplitude oscillations such as wave leakage from the prominences, but may also be caused by non-linear effects \citep[see ][and references therein]{oliver_2009}.

\section{Trigger of Oscillation\label{trigger}}

In the earlier observations, for example by \cite{ramsey_smith_1966}, and more recently by \cite{eto_2002, gilbert_2008} all the reported oscillations were caused by the interaction of the filament with the Moreton wave \citep[][]{moreton}. However, some recent observations clearly suggest that large amplitude oscillations could be present in prominences without the existence of a remote flare and the associated wave phenomena. They may be caused by nearby sub-flares
or jets \citep[see e.g., ][]{jing_2003, jing_2006, isobeetal_2007, vrsnak_2007}.

\begin{figure}
\centering
\includegraphics[width=1.0\textwidth]{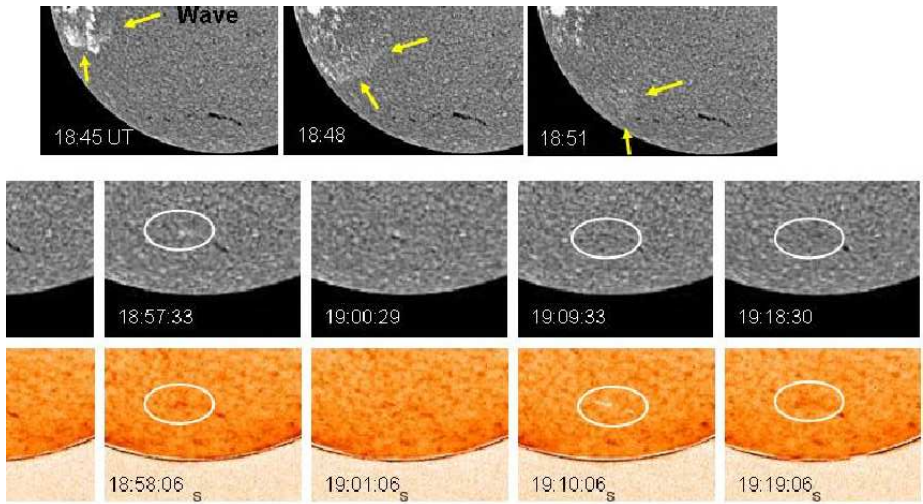}
\caption{Wave observed in H$\alpha$ data from MLSO (top) and the initial stages of response to the passing
wave in H$\alpha$ (middle) and He I ($\lambda$10830) intensity (bottom) from MLSO. White circles show
the largest differences in appearance in the filament in the two lines.\label{gilbert_1}}
\end{figure}

Despite the fact that there have been several observations of filament oscillations due to interaction with waves generated by solar flares, the exact nature of relationship between the properties of the wave (i.e., the speed, energy, and topology) and the filament activation is currently not well understood. \cite{gilbert_2008} provided a comprehensive study of a wave-filament interaction using chromospheric observations recorded at the Mauna Loa Solar Observatory (MLSO). The MLSO detected a Moreton wave associated with a flare on 2006 December 6, which interacted with a filament producing a large amplitude oscillation in the filament with an amplitude of about 41~km~s$^{-1}$ and a period of $\approx$29~mins. Fig.~\ref{gilbert_1} shows the wave propagation in H$\alpha$ (top
row) and the initial stages of the filament activation in
co-temporal H$\alpha$ (middle row) and He~I~($\lambda$10830)
intensity (bottom row) data. There is a slight difference in the
appearance of the filament in the two lines at 18:57:33, 19:09:33,
and 19:18:30 UT (H$\alpha$ times), marked with white circles in the
figure. The filament's disappearance in H$\alpha$ represents what
has historically been referred to as "winking," and occurs because
at large velocities ($\approx$30-40~km~s$^{-1}$ for Polarimeter for Inner Coronal Studies (PICS) , depending on how dark the original filament is), the filament material will shift out of the
narrow pass-band. Using a filament mass of about
4$\times$10$^{14}$ g and the inferred maximum LOS
velocity of 41~km~s$^{-1}$ (red shift) and 21~km~s$^{-1}$ (blue
shift), \cite{gilbert_2008} estimated the kinetic energy of the
order of 10$^{20}$ Joule and 10$^{19}$ Joule in the red and blue
wing respectively. This is in agreement with the predicted energy
required to induce quiescent prominence oscillations as 10$^{19}$
- 10$^{20}$ Joules, as provided by \cite{k_k}.

An interesting observation was reported by \cite{okamoto_2004} in which the large amplitude oscillation was excited by the interaction with an EIT wave. The EIT waves were first discovered by \cite{thompson_1998} in the observations recorded by the EIT instrument aboard SoHO and hence the name. Using H$\alpha$ center and $\pm$0.8~{\AA} full disk images taken by the Flare Monitoring Telescope (FMT) at Hida Observatory \citep{kurokawa_1995}, \cite{okamoto_2004} found four winking
filaments spread over the solar disk, apparently triggered by a
single X-class flare on 10 Apr 2001. In this event no associated
Moreton wave was detected in FMT observations, but a type II radio
burst was found in the Hiraiso Radio Spectrograph
\citep[HiRAS;][]{kondo} observations, indicating the presence of a
coronal MHD shock wave. \cite{okamoto_2004} compared the
propagation speed of the EIT waves and the type II radio bursts
with the timings of the onset of the filament oscillations. They
concluded that (1) the propagation speed of the EIT wave was
different from that of the type-II burst, indicating their
different origin, and (2) at least three out of four oscillations
were triggered by the EIT wave and not by the fast-mode coronal
shock, i.e., the type II radio burst.

As mentioned earlier, the name of "EIT wave" was coined because
such large-scale coronal transient were commonly observed by the
SOHO/EIT. However similar wave-like phenomena have been found in
other EUV instruments, such as TRACE and STEREO. Therefore it is
more appropriate to call them "EUV waves". Although referred to as
a ``\textit{wave}'', whether the EUV waves  are fast-mode MHD
waves or not, is still debatable \citep[see e.g.,][]{chen_shibata, attrill, t_r, delannee,  p_v}. The fact that the
EUV waves were able to initiate large amplitude oscillations in
filaments may provide some insight into their nature. Further,
observations as well as theoretical and numerical studies of the
interaction of a filament and the EUV waves are desired.

Three out of four filament oscillations presented by \cite{jing_2003, jing_2006} were active region filaments, while one was a quiescent filament. For the active region filaments, the oscillations were associated with either sub-flares or c-class flares. However, there were no associated activity observed with the quiescent filament. The observation presented by \cite{isobe_tripathi_2006} suggest that it was rather associated
with a ``jet-like phenomenon'' at the time of the onset of the oscillation. For the same event \cite{isobeetal_2007} found that the jet was associated with an emerging photospheric flux and a small scale brightening seen in EIT images. Based on these observational facts, it was concluded that this oscillation seems to have been triggered due to magnetic reconnection between a nearby emerging magnetic flux and a filament barb \citep[see][for more details]{isobeetal_2007}. \cite{vrsnak_2007}  presented observations of an oscillating filament which was associated with a flare-like brightening just before or around the onset of the oscillation, providing evidence of magnetic reconnection. However,
there were no associated activity, such as flux emergence or cancellation, observed for the event presented by \cite{vrsnak_2007}.  It is worthwhile emphasizing that, for all the observations discussed by \cite{jing_2003, jing_2006, isobe_tripathi_2006, vrsnak_2007} there were no obvious disturbances such as Moreton waves or EUV waves produced by a remote flare triggering these oscillations.

\section{Prominence seismology in large amplitude oscillation}\label{seismology}

The fact that filaments have their own characteristic frequency as suggested by the observations (see \cite{ramsey_smith_1966}) clearly demonstrated that a filament can oscillate with its global eigen mode and eigen frequency. This characterstic makes it a strong candidate for comparison with theoretical MHD wave models.

It is due to substantial developments in observational and data-analysis techniques as well as theory of prominence oscillations over the years, considerable improvements have been
made in the field of prominence seismology \citep[for a review see e.g.,][and references therein]{oliver_review, oliver_2009}. The
study of prominence oscillations provides an alternative approach to probing their internal magnetic field configuration, physical
plasma parameters, and their stability in the solar corona. These
are rather difficult otherwise. Combining observations and
theories of prominence oscillations is proving invaluable in
probing the internal magnetic field structure and other physical
plasma properties in prominences, justifying the field of
prominence seismology as a powerful diagnostic tool.

One of the first applications of prominence seismology was performed by \cite{hyder_1966}. \cite{hyder_1966} applied the \cite{k_s} model of prominences to the winking filaments reported by \cite{ramsey_smith_1966}. In this work, \cite{hyder_1966} considered the filament oscillations as a damped harmonic oscillator and magnetic tension as the restoring force and estimated the radial component of the magnetic field in the filament to be between 2 and 30 Gauss and the coronal component of viscosity between 4~$\times 10^{-10}$ and 1.6~$\times 10^{-9}$ poise. The values derived using this method were in agreement with those obtained with the magnetograph measurements of \cite{zirin} and other methods such as analysis of the polarization of prominence H$\alpha$ and D3 and with the prominence magnetograph measurements by \cite{lee}. The study of \cite{hyder_1966} clearly demonstrated that prominence seismology is an excellent tool for deriving different physical parameters such as internal magnetic field which are often very complicated by other means.

\cite{vrsnak_2007} applied the idea of prominence seismology to a prominence oscillating in longitudinal direction recorded on 23-Jan-2002 in H$\alpha$ observations with a period of 50 minutes. They believe that this oscillation was triggered by the associated small scale brightening as a consequence of magnetic reconnetion. Based on the assumption that the magnetic reconnetion would inject a certain amount of poloidal flux in the filament and the filament is representative of a flux rope, they developed a simple model to deduce the poloidal and axial component of the magentic field, considering the oscillation to be a simple harmonic motion and that it was oscillating in longitudinal direction. Using this approach they derived the axial component of magnetic field to be 10-30 Gauss, which is reasonable for a quiescent prominences.

\begin{figure}
\centering
\includegraphics[width=1.0\textwidth]{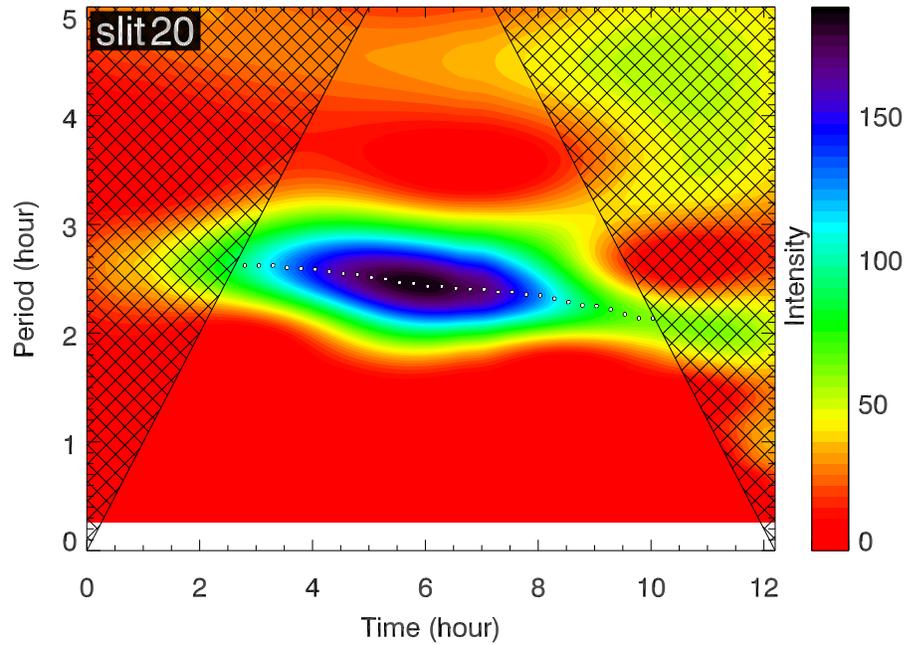}
\caption{Wavelet spectrum of the filament position for a specific location on the oscillating prominence. The dotted line indicates the average of the dominant periods for each time-snapshot shown on the x-axis. See \cite{pinter} for detailed version of this figure. \label{pinter_fig}}
\end{figure}

Recently, \cite{pinter} took the idea of Prominence seismology further by investigating the temporal and spatial behaviour of a large amplitude filament oscillation seen in a polar crown filament using wavelet analysis. The oscillations were first reported by \cite{isobe_tripathi_2006} who analyzed the phenomena using images taken by the EIT. They found that the oscillation repeated at least three times before the filament erupted. As further investigated by \cite{isobeetal_2007}, the amplitudes of velocity and spatial displacement of the oscillation in the POS were about 5 km/s and 15000 km, respectively. \cite{pinter} investigated, in detail, the temporal and spatial variation of the intensity distributions along and across the filament to study the filament oscillation. They put forward two different methods to follow the filament motion by dividing the entire filament into 49 equally spaced parallel slits: (a) "global fitting" method where a quadratic polynomial was fitted to the two-dimensional intensity distribution of the filament region, (b) "minimum search" method where the filament position was represented by the position of the local minimum in the intensity distribution along each slit at every time snapshot. The wavelet analysis revealed similar results.  For each method, the period of oscillation was determined to be 2.5 to 2.6 hour which seemed to decrease slightly toward the center of the filament (see Fig.~\ref{pinter_fig}). They also found that the oscillation was more dominant around the middle of the filament with no significant motions detected at the two endpoints. We note here that only a part of the polar crown filament was oscillating. By investigating the positive and negative peaks in each slit, \cite{pinter} concluded that the "global" oscillation of the filament were transverse to the filament axis. Since the helical structure of the filament threads are clearly visible in the images, the filament appears to have a twisted flux-rope like structure and thus using the model of \cite{vrsnak_2007} for a twisted flux-rope, in which the flux-rope was essentially considered to be a simple harmoic oscillator, \cite{pinter} estimated the poloidal and axial components of magnetic field in the filament to be between 2-10 Gauss and 1-5 Gauss respectively. It is worthwhile mentioning that the sense of oscillation observed by \cite{pinter} was trasverse and the model developed by \cite{vrsnak_2007} was for longitudinal oscillations as described above. However, for the same event as studied by \cite{pinter}, \cite{isobe_tripathi_2006} applied a simple model developed by \cite{k_k} for a freely oscillating prominence and found similar results. We note here that, although we have got similar results by using two different simple models of filament oscillations, we are still not in position to deduce any kind of internal magnetic structure of the filament using the observationally derived parameteres. Therefore, further advancements, both theoretically and observationally, are in order.

\section{Prominence seismology as a diagnostic tool for stability and eruption mechanism of oscillating prominences}\label{eruption}
A prominence can accommodate large-amplitude oscillations when it
is in an equilibrium that is stable against the large-amplitude
disturbance. Therefore, prominence seismology can be used not only
for determining the physical parameters but also for diagnosing
the stability of the prominence. In particular, when a
large-amplitude oscillation occurs while the prominence is
erupting or is close to eruption, it may provide some information
on the onset mechanism of the eruption. Unfortunately, large
amplitude oscillations are rare events. The probability of
observing such a large amplitude oscillation during eruption is
very low.

\begin{figure}
\centering
\includegraphics[width=1.0\textwidth]{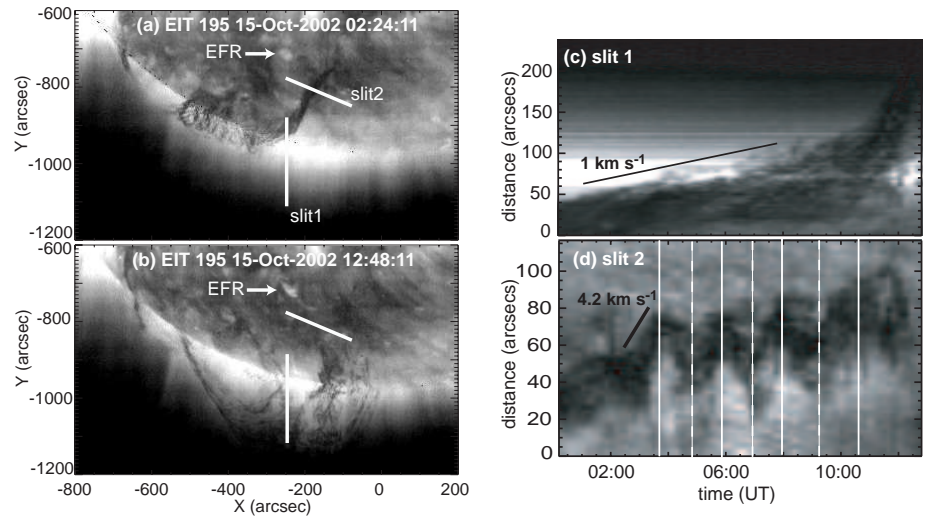}
\caption{Prominence oscillation event on 15 Oct 2002 (adopted from \cite{isobe_tripathi_2006}). (a-b): EIT images of the prominence during it slow-rise phase (a) and erupting phase (b) An nearby emerging flux is indicated as EFR. (c-d): Time slices of the slit 1(a) and slit (2) indicated in panels (a) and (b).\label{15oct02}}
\end{figure}

To the best of our knowledge the 15 Oct 2002 event first reported by \cite{isobe_tripathi_2006}
and subsequently studied in detail by \cite{isobeetal_2007} and \cite{pinter}, is the only event reported in the literature where large-amplitude oscillation occurred during eruption. As shown in Fig.~\ref{15oct02}, the oscillation occurred while the prominence is slowly rising with an apparent LOS velocity of 1~km~s$^{-1}$. The oscillation completed three cycles, and during the forth cycle the prominence was suddenly accelerated and erupted. The slow-rise phase in prominences before the fast eruption is commonly observed phenomena  \citep{sterling_moore_2005, chifor_2006, f_w, chifor_2007, nagashima_2007}.

This observation has two significant implications for the onset of eruption. On one hand, the existence of oscillation in the slow-rise phase is the evidence that the prominence, or at least its erupting portion, retains non-linearly stable equilibrium during the slow rise. Therefore, it is neither a slowly growing phase of an instability nor the dynamic motion after the prominence lost it equilibrium. Rather, it must be a quasi-static evolution. On the other hand, the time scale of the transition from such non-linearly stable equilibrium to the fast eruption (i.e., instability or loss of equilibrium) occurred on a time scale shorter than the period of oscillation of the prominence. If the eruption occurred through the exchange of stability driven by a shearing/converging motion in the photosphere, one expects that the period of oscillation becomes longer as the restoring force becomes weaker. However, according to the wavelet analysis by \cite{pinter}, the period of oscillation during the 3-4 cycles did not increase and indeed showed a slight decrease. Therefore, the "trigger" for the transition from slow-rise to fast eruption should be a fast mechanism such as magnetic reconnection.

We should note, however, that in this event only a portion of the prominence exhibited oscillation, and rest of the prominence did not show any signature of oscillation during the slow-rise phase (see the slit 1 in Fig. \ref{15oct02} and \cite{isobeetal_2007} for more detail). Hence it is possible that the oscillating part is locally in a nonlinearly stable equilibrium, but the other part of the filament is already unstable. If this is the case, the transition from stable equilibrium to eruption of the {\it oscillating part} may be triggered by the rise of the other (already unstable) part of the prominence, as suggested by \cite{tripathi, chifor_2006, chifor_2007}. In conclusion, prominence seismology is potentially a powerful tool to diagnose the eruption mechanisms. See also discussions in \cite{foullon, foullon_2009} and \cite{f_w}. Further observations as well as detailed theoretical investigations are desired.

\section{Summary and Discussion} \label{summary}

Large amplitude oscillations in filaments (prominences) are rare events. There have been $\sim 20$ events reported in the literature so far. In this paper we have tried to provide a comprehensive observational overview on large amplitude oscillations of prominences based on the research published so far. The origin of the large amplitude oscillations is not fully understood but it is believed that a filament exhibits large-amplitude oscillation when it undergoes large-amplitude disturbance by the coronal shock waves from a distant flare or from nearby sub-flares or jets. When the oscillations were initiated by a shock wave (e.g., Moreton wave) from distant flare, the filament oscillates almost as a solid body. On the other hand, when a small event in the vicinity of the filament is the trigger of the oscillation, only a portion of the filament oscillates as found by \cite{jing_2003, jing_2006, isobe_tripathi_2006, vrsnak_2007}.

In cases where, the large amplitude oscillations are due to the waves produced by a flare, the exact relationship between the wave's physical parameter and that of the filament oscillation is not fully understood. After studying oscillations in 11 filaments, \cite{ramsey_smith_1966} pointed out that the filaments have their own characteristic frequencies of oscillation which are independent of the size of the flare, the distance from the flare and the inferred wave velocity of the propagating disturbances originated from the flare site. But the direction of motion and the amplitude seems closely related to the characteristics of the disturbance. However, currently it is not clear what the oscillation frequency depends on. We know that filaments are cool chromospheric plasma suspended in the solar atmosphere supported by magnetic field. Therefore, the geometry and the strength of magnetic field in prominences should play a major role in determining the characteristic frequency of oscillation in prominences. This clearly indicates the crucial role the subject of prominence seismology can play in making further contributions in unraveling the internal structure of prominences.

Prominence seismology is now applied to diagnose the  magnetic field strength whose direct measurement is difficult. Moreover, it is potentially very powerful tool to diagnose the stability and the eruption mechanisms, as was demonstrated by \cite{isobeetal_2007}. The occurrence of large amplitude
oscillations initiated by some external trigger in an erupting prominence seem to be very rare, but it is worth trying to search for more such events.

We have mentioned that some oscillating filaments erupt after a few cycles whereas others don't, suggesting a damping mechanism at play in the oscillations. However, due to the small number of observed events, the reason of this damping is not quite well understood. In some cases this can be attributed to an aerodynamic drag or some dissipative mechanisms, such as radiative losses, viscous damping, wave leakage and ion-neutral collisions as discussed by various authors. However, different damping time scales produced by different mechanisms require a proper comparison with the observations.

It is noteworthy that in order to derive the oscillatory parameters the position and/or the Doppler velocity of a single position on the filament is commonly used. This is good enough when the filaments oscillates as a whole i.e., every single point along the filament moves in phase. However such analysis can lead
to wrong results if the whole filament is not moving in phase. It
has been observed that the whole filament moves in phase when the
oscillation is triggered by  a wave generated at a remote flare
site. On the other hand it has also been observed that only a part
of the prominence oscillates. This usually happens when the
oscillatory motion is associated with a sub-flare or micro-flare
for example. Therefore, it would be important to investigate the signal
from many points along the oscillating prominence and to ensure
that all the points along the prominence are in phase. In
addition, the prominence oscillations which are damped in time,
are fitted with damped sinusoidal signals in order to obtain the
oscillation period and the damping time. Other tools such as the
power spectrum or the wavelet analysis have been scarcely used. We
understand that there are a large number of non-linear effects in
oscillating prominences. These may modify the physical properties
of prominences during the oscillations, which implies that the
oscillation period and damping time may vary in time. This
possible variation could be revealed by the powerful wavelet
diagram. So far, to the best of our knowledge, in large amplitude
oscillation events, this kind of analysis is only performed by
\cite{pinter}.

In future, it will be very interesting to observe large amplitude
oscillations with high spatial/temporal resolution. It
would be worthwhile to search for large amplitude oscillation
events which oscillates both in transverse and longitudinal
direction and the ones that are triggered by different mechanisms.
This would allow us to make definite conclusions about the
dependence of the periods on the direction as well as the trigger
mechanisms. Observations by the Solar Optical Telescope
\citep{tsuneta_2008} on board Hinode have shown striking variety
of small scale dynamics such as oscillating threads
\citep{okamoto_2007}, rising  rising dark plume-like structure
\citep{berger_2008}. If one can observe large amplitude
disturbances with such high resolution, it will provide
significant information on the internal structure of the
prominence and the nature of these small scale dynamics. In
addition, if prominence oscillations are observed by the Solar
Terrestrial Relations Observatory (STEREO), it will provide an
unambiguous interpretation of the direction of motion.

\begin{acknowledgements} We thank the two independent referees for their thoughtful comments which improved the manuscript. DT acknowledges the support from STFC and ISSI to provide funds to
travel to attend the workshop in Bern. HI acknowdledge the support
from the Grant-in-Aid for Creative Scientific Researchâ The Basic
Study of Space Weather Prediction" from MEXT, Japan (Head
Investigator: K. Shibata). RJ would like to thank Dr. B. Pint\'er.
She is also indebted to Prof. B. Roberts for developing her
interest over the years in the field of MHD waves and
oscillations.
\end{acknowledgements}

\bibliography{references.bib}   

\begin{landscape}
\begin{table}[]
\caption{Physical parameters of large amplitude oscillations reported so far in the literature. In the table from left to right: date and type of filament in which the oscillations were observed, oscillations amplitude (km s$^{-1}$), period (minutes), damping time (minutes), direction of motion, trigger of oscillation, kind of data in which the oscillations was observed and the papers where these oscillations were reported.\label{table}}
\begin{center}
\begin{tabular*}{18cm}{c|lllllll} \hline
    Date                        & Amplitude             & Period            & Damping           &  Direction    &   Trigger                 &   Data            &  Paper    \\
                                & (km s$^{-1}$)         & (min)             & Time (min)\\
        \hline

        11 events               &$-$                    & 6$-$40                & $-$                   & $-$                   & $-$                       & $-$               & \cite{ramsey_smith_1966}\\
        (Before 1966)\\\\

        4 Nov 1997              & $-$                   &  15                       & $-$                   &  Transverse, vertical             &   Moreton wave    &   H$\alpha$ wink  & \cite{eto_2002}    \\
        (QS filament)\\ \\

        24 Oct 2001             & 92                    & 80                        & 210                   &  Longitudinal         & Nearby sub-flare  &   H$\alpha$       & \cite{jing_2003}   \\
        (AR filament)\\\\

        10 Apr 2001             & $-$                   &  28                       & $-$                   &  Transverse, vertical         & EIT wave              &   H$\alpha$ wink  & \cite{okamoto_2004} \\
        (QS filament)\\\\

        24 Oct 2001                    &50                          & 160                              & 600                                & Longitudinal                        & C-class flare       & H$\alpha$     & \cite{jing_2006}   \\
        (AR filament) \\\\

    20 March 2002       &30         & 150               &$-$                & Longitudinal      &near by micro-flare    &H$\alpha$  & \cite{jing_2006}\\
    (AR filament)\\\\

    22 March 2002       &100            & 100               &$-$                &   Longitudinal        &  $-$              & H$\alpha$     & \cite{jing_2006}\\
    (QS filament)\\\\

        15 Oct 2002             & 20                &  150                  & $-$                   & Transverse, horizontal& Nearby jet        & EIT, H$\alpha$ wink & \cite{isobe_tripathi_2006} \\
        (Polar crown\&  &                       &                           &                           &                       &                   & Radio             &   \cite{isobeetal_2007}\\
            eruption)           &                       &                           &                           &                       &                   &                   &   \cite{pinter}\\\\

        23 Jan 2002     & 51            & 50            & 115       & Longitudinal     & Nearby sub-flare  & H$\alpha$         & \cite{vrsnak_2007}    \\
        (AR filament)\\\\

        06 Dec 2006     & 41            & 29            & 180       & Transverse vertical      & Moreton wave      & He 10830          & \cite{gilbert_2008}   \\
        (QS filament)\\\\
        \end{tabular*}
      \end{center}
    \end{table}
    \end{landscape}

\end{document}